\documentclass[prl,twocolumn,showpacs,superscriptaddress]{revtex4}
\usepackage{graphicx}
\usepackage{amsmath,amssymb}
\usepackage{color}

\newcommand{\myav}[1]{\langle #1\rangle}

\newcommand{\mysection}[1]{\subsection{#1}}

\newcommand{\myvec}[1]{{\vec{#1}}}
\newcommand{\rvec}{\myvec{r}}
\newcommand{\fvec}{\myvec{f}}
\newcommand{\etavec}{\myvec{\eta}}
\newcommand{\xivec}{\myvec{\xi}}
\newcommand{\vvec}{\myvec{v}}

\newcommand{\kB}{k_\mathrm{B}}
\newcommand{\kT}{\kB T}
\newcommand{\Peq}{P_{\mathrm{eq}}}
\newcommand{\Pss}{P_{\mathrm{ss}}}
\newcommand{\Teff}{T_{\mathrm{eff}}}
\newcommand{\dt}{\delta t}
\newcommand{\Dt}{\Delta t}
\newcommand{\gdot}{\dot\gamma}
\newcommand{\ddt}[1]{\frac{d{#1}}{dt}}

\newcommand{\delr}{\Delta_i}
\newcommand{\delq}{\Delta_q}

\newcommand{\latin}[1]{{\itshape #1}}
\newcommand{\eg}{\latin{e.\,g.}}
\newcommand{\ie}{\latin{i.\,e.}}
\newcommand{\etal}{\latin{et al.}}
\newcommand{\via}{\latin{via}}
\newcommand{\viceversa}{\latin{vice versa}}
\newcommand{\cf}{\latin{cf.}}
\newcommand{\ansatz}{\emph{ansatz}}
\newcommand{\letter}{Letter}

\begin{document}

\title{Computing sensitivity coefficients in Brownian dynamics
  simulations by Malliavin weight sampling}

\author{Patrick B. Warren}

\affiliation{Unilever R\&D Port Sunlight, Quarry Road East, Bebington,
  Wirral, CH63 3JW, UK.}

\author{Rosalind J. Allen}

\affiliation{SUPA, School of Physics and Astronomy, The University of
  Edinburgh, The Kings Buildings, Mayfield Road, Edinburgh, EH9 3JZ,
  UK.}

\date{submitted version --- July 11, 2012}

\begin{abstract}
We present a method for computing parameter sensitivities and response
coefficients in Brownian dynamics simulations. The method involves
tracking auxiliary variables (Malliavin weights) in addition to the
usual particle positions, in an unperturbed simulation.  The Malliavin
weights sample the derivatives of the probability density with respect
to the parameters of interest and are also interesting dynamical
objects in themselves.  Malliavin weight sampling is simple to
implement, applies to equilibrium or nonequilibrium, steady state or
time-dependent systems, and scales more efficiently than standard
finite difference methods.
\end{abstract}

 %  Nonequilibrium and irreversible thermodynamics

\maketitle

The response of a system to infinitesimal changes in an external field
provides important insights into its underlying physics. Divergences
in such response functions can indicate the presence of phase
transitions, while their relation to fluctuations in unperturbed
systems \via\ fluctuation-dissipation theorems (FDTs) provides a key
diagnostic of the difference between equilibrium and non-equilibrium
systems \cite{CL95}. Knowledge of the response of a system to changes
in \emph{internal} parameters (\eg\ those controlling inter-particle
forces) is also of great importance, since it has the potential
greatly to accelerate the fitting of force fields to experimental
data. For equilibrium systems, responses to perturbations can often be
computed from the properties of unperturbed systems via known
statistical mechanical relations \cite{HT82, CKP94, PJP09}. However,
for systems which are far from steady-state, and/or whose dynamics
does not obey detailed balance, one is generally forced to resort to
finite differencing: explicitly taking the difference between
simulation trajectories generated at slightly different parameter
values \cite{CJ75}. While finite differencing can be made more
efficient by reuse of random number streams \cite{LE92, RSK10}, one
still has to re-simulate the perturbed system for each parameter of
interest.

In this \letter, we present a simple and generic method for computing
responses to infinitesimal changes in internal or external parameters
in stochastic Brownian dynamics simulations, which may be in or out of
equilibrium.  The method does not require simulation of the perturbed
system; instead, it involves tracking, in an unperturbed system,
auxiliary stochastic variables which sample the derivatives of the
probability density with respect to the parameters of interest. We
term these auxiliary variables `Malliavin weights' as the method has
close links to the Malliavin calculus programme \cite{Nua06-Bel07},
used in quantitative finance for deriving price sensitivities (\ie\
`Greeks') \cite{FLL+99-CG07}. Our method extends approaches previously
proposed for kinetic Monte-Carlo simulations \cite{Ber07, PA07, WA12}
to a much wider set of problems. It also has interesting links to
molecular dynamics methods in which response coefficients are computed
via the integration of adjunct equations of motion for individual
particles \cite{CJ75, Ber07}.  Since Brownian dynamics is very widely
used \cite{EM78-AT87} in the study of non-equilibrium statistical
physics problems such as driven steady states, active soft matter, and
modeling sub-cellular processes in biology, we anticipate that our
method should prove widely applicable.
  
We begin by considering a collection of $i=1\dots N$ interacting
particles undergoing overdamped Brownian motion, described by the
coupled Langevin equations
\begin{equation}
\ddt{\rvec_i}=\frac{D\fvec_i}{\kT}  + \etavec_i\,.
\label{eq:gle}
\end{equation}
Here $\rvec_i$ and $\fvec_i$ are the position of the $i$th particle
and the force acting on it, respectively, $D$ is the diffusion
coefficient (which for simplicity we here assume to be constant and the
same for all particles), $\kB$ is Boltzmann's constant, $T$ is
temperature, and the $\etavec_i$ are independent vectors of Gaussian
white noise of amplitude $2D$.  We now add an extra variable
$q_\lambda$ (a Malliavin weight) which evolves according to
\begin{equation}
\ddt{q_\lambda}=\frac{1}{2\kT} \sum_{i=1}^N
\frac{\partial\fvec_i}{\partial\lambda}\cdot\etavec_i
\label{eq:gmw}
\end{equation}
where $\lambda$ is a parameter of interest for which the only
requirement is that $\partial \fvec_i / \partial \lambda$ is known.
Note that $q_\lambda$ does not perturb the dynamics of the particles;
it merely acts as a `readout' and should be initialised to
$q_\lambda=0$.  The interpretation of Eq.~\eqref{eq:gmw} as a
stochastic differential equation is straightforward and uniquely
defined---the practical implementation is described in Supplementary
Material \cite{suppl-note}.  The noise vector $\etavec_i$ is identical
in Eqs.~\eqref{eq:gle} and \eqref{eq:gmw}---in each Brownian dynamics
timestep, $q_\lambda$ is updated using the \emph{same set of random
  numbers} that were chosen in the update of the particle positions.
Our central claim is that for any function of the particle positions
$A(\{\rvec_i\})$
\begin{equation}
\frac{\partial\myav{A}}{\partial\lambda} = \myav{A\,q_\lambda}\,,
\label{eq:mall}
\end{equation}
\ie\ the response of $A$ to the parameter $\lambda$ is given by the
average of $A$ in the unperturbed system, weighted by the appropriate
Malliavin weight $q_\lambda$.  Eq.~\eqref{eq:mall} has important
practical implications.  Since the computation of $q_\lambda$ via
Eq.~\eqref{eq:gmw} is independent of $A$, the same $q_\lambda$ can be
used to compute the sensitivity of multiple system properties to the
parameter $\lambda$.  Moreover, one can track multiple weights
corresponding to different choices of $\lambda$, with marginal
additional cost.

Eq.~\eqref{eq:mall} is the key result of this Letter.  It can be
proved by taking moments of a Chapman-Kolmogorov equation for the
evolution of the \emph{joint} probability distribution
$P(\{\rvec_i\},q_\lambda;t)$ for the set of particle positions
$\{\rvec_i\}$ and the Malliavin weight $q_\lambda$.  The details are
given in Supplementary Material \cite{suppl-note, alt-note}.  A
crucial intermediate result is that the conditional average of
$q_\lambda$ for a given set of particle positions $\{\rvec_i\}$, which
we denote $\myav{q_\lambda}_{\{\rvec_i\}}$, is given by
\begin{equation}
\myav{q_\lambda}_{\{\rvec_i\}}\equiv 
\frac{\int\!dq_\lambda\,q_\lambda\,P(\{\rvec_i\},q_\lambda;t)}
{\int\!dq_\lambda\,P(\{\rvec_i\},q_\lambda;t)} = 
\frac{\partial \ln P(\{\rvec_i\};t)}{\partial \lambda}
\label{eq:condav}
\end{equation}
where $P(\{\rvec_i\};t)$ is the probability distribution for the
particle positions.  Thus the Malliavin weight in fact samples the
conjugate \cite{HT82} variable $\partial\ln P/\partial\lambda$.

The proof makes no assumptions about the system being in steady state
or obeying detailed balance; thus our method is valid for systems far
from steady state, or in driven steady states, as well as for those at
equilibrium. Moreover, as we show in Supplementary Material
\cite{suppl-note}, our approach can easily be extended to systems
undergoing underdamped Brownian motion and to the computation of
higher-order derivatives.

\begin{figure}
\begin{center}
\includegraphics[clip=true,width=0.8\columnwidth]{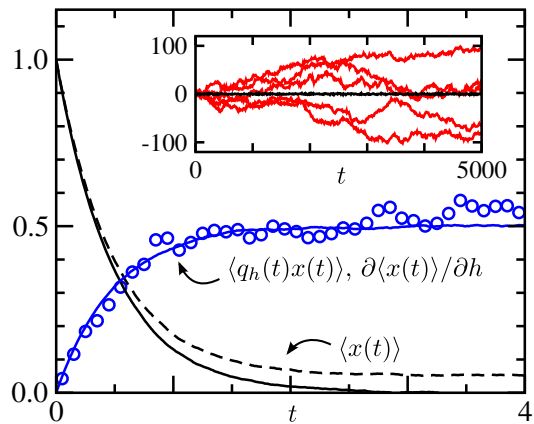}
\end{center}
\vskip -0.5cm
\caption{(color online) Numerical simulation of Eqs.~\eqref{eq:1trap}
  with $\kappa = 2$, $x_0=1$, and $D=T=1$.  The falling curves show
  $\myav{x(t)}$ for $h=0$ (solid line) and $h=0.1$ (dashed line).  The
  rising curves show $\myav{x(t)q_h(t)}$ (blue solid line) and
  $\partial\myav{x(t)}/\partial h$ evaluated using forward finite
  differencing (blue circles).  Averages are over $10^5$ replicate
  simulations.  The inset shows the long-time behaviour of individual
  $x(t)$ and $q_h(t)$ trajectories.  While $x(t)$ stays close to its
  equilibrium value (black), $q_h(t)$ (five replicates, red) executes
  a random walk.}
\label{fig:trap1d}
\end{figure}

To illustrate the method, we turn to a simple example for which
analytical results are available: a single particle in a
one-dimensional harmonic trap described by a potential
$U=\frac{1}{2}\kappa x^2-hx$.  To make contact with linear response
theory, we take the parameter of interest to be the strength of
the applied external force $h$.  We set the particle mobility to unity
so that the diffusion constant $D=T$ where the temperature $T$ is in
units of $\kB$. At equilibrium, $\Peq(x)\sim e^{-U/T}$, so that
$\partial\ln\Peq/\partial h = (x-\myav{x})/T$ \cite{PJP09}, and
the FDT holds: $\partial\myav{x}/\partial h =
(\myav{x^2}-\myav{x}^2)/T$. We now apply Malliavin weight sampling to
the time-dependent situation in which the particle starts from
$x=x_0$ at $t=0$ and relaxes towards its equilibrium
position. Eqs.~\eqref{eq:gle} and \eqref{eq:gmw} become
\begin{equation}
\ddt{x}=-\kappa x+ h+\eta,\quad
\ddt{q_h}=\frac{\eta}{2T}\,.\label{eq:1trap}
\end{equation}
These equations can be solved exactly \cite{suppl-note} to give
$P(x,q_h; t)$ as a bivariate Gaussian with
\begin{equation}
\begin{array}{l}
%\displaystyle
\myav{x}=x_0 e^{-\kappa t}+({h}/{\kappa})(1-e^{-\kappa
    t})\,,\quad\myav{q_h}=0\,,\\[6pt]
%\displaystyle
\myav{x^2}-\myav{x}^2=
({T}/{\kappa})(1-e^{-2\kappa t})\,,
\\[6pt]
%\displaystyle
\myav{x q_h}=({1-e^{-\kappa t}})/{\kappa}\,,\quad 
\myav{q_h^2}={t}/{2T}\,.
\end{array}\label{eq:1dtrap}
\end{equation}
This result allows us to verify directly that Eq.~\eqref{eq:condav} is
satisfied, that is to say $\myav{q_h}_x = \int\! dx\,q_h P(x,q_h;t) =
\partial\ln P(x;t)/\partial h$, where $P(x;t)=\int\!
dq_h\,P(x,q_h;t)$. Fig.~\ref{fig:trap1d} shows simulation results for
$\partial \myav{x}/\partial h$, computed by Malliavin weight sampling
(MWS) and by forward finite differencing; the inset shows trajectories
for $q_h$ from replicate simulation runs. The Malliavin weight $q_h$
behaves as a random walk with zero mean and a diffusion coefficient
$1/4T$.  Eq.~\eqref{eq:1dtrap} shows that, by analogy with the
equilibrium FDT, one can define a time-dependent effective temperature
$\Teff/T = (1-e^{-2\kappa t})/(1-e^{-\kappa t})$ such that
$\partial\myav{x}/\partial h = (\myav{x^2}-\myav{x}^2)/\Teff$.
Calculating the conditional average of the Malliavin weight gives
$\myav{q_h}_x = (x-\myav{x})/\Teff$.  Interestingly this has the same
form as the equilibrium case but again features $\Teff$.  Thus the
effective temperature has a wider relevance than would be apparent
from the time-dependent FDT since $\partial\myav{A}/\partial h =
(\myav{Ax}-\myav{A}\myav{x})/\Teff$, where $A(x)$ is any function of
the particle position.

\begin{figure}
\begin{center}
\includegraphics[clip=true,width=0.8\columnwidth]{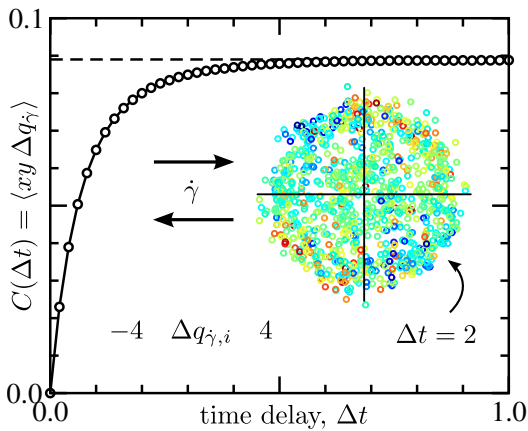}
\end{center}
\vskip -0.5cm
\caption{(color online) Interacting particles in a two-dimensional
  harmonic trap under shear. The main plot shows how the correlation
  function $\myav{x y \Delta q_{\gdot}}$ (solid line; calculated in an
  unsheared system) asymptotes to
  $\partial\myav{xy}/\partial\gdot|_{\gdot=0} = 0.0890 \pm 0.0005$
  calculated by centered finite differencing (dashed line).  The inset
  shows 100 superimposed simulation snapshots at $\Dt=2$, colored by
  individual particle Malliavin weight.}
\label{fig:trap25}
\end{figure}

An important practical issue is raised by the fact that $q_\lambda(t)$
behaves as a random walk (inset to Figure \ref{fig:trap1d}): to
compute responses to parameter perturbations for systems in steady
state, we cannot simply monitor $q_\lambda$ for longer and longer
times until the system reaches its steady state. This is because
replicate trajectories of $q_\lambda$ diverge at long times and
measurements of $\myav{A(t) q_\lambda(t)}$ incur a large sampling
error. Fortunately, we can circumvent this problem by computing
instead the correlation function \cite{Ber07, WA12} $C(t, t_0)=\myav{
  A(t)\,\Delta q_\lambda(t,t_0)}$ where $\Delta q_\lambda(t,t_0)=
q_\lambda(t)-q_\lambda(t_0)$.  In steady state this becomes a
well-defined function of $\Dt=t-t_0$ and we expect it to obey
$C(\Delta t)\to\partial\myav{A}/\partial\lambda$ as $\Dt\to\infty$
\cite{WA12, suppl-note}.  Like any other correlation function, $C(t)$
converges to its asymptotic value on a time scale set by the spectrum
of relaxation times in the problem.

Next, we demonstrate the application of MWS to a much less trivial
example: a non-equilibrium driven steady state formed by a cluster of
particles in a two-dimensional harmonic trap, under shear. This
example is motivated by recent experimental studies of colloidal
particles in optical traps, which have provided new insights into
statistical physics at the microscale \cite{KSG02-KS11}. We suppose
that the particles interact with each other via a repulsive screened
Coulomb potential $u(r)=(\Gamma/r) e^{-r/r_c}$, with coupling strength
$\Gamma$ and range $r_c$. The total potential energy of the cluster is
then $U=\sum_{i>j} u(r_{ij})+\frac{1}{2}\kappa \sum_i r_i^2$ (where
$\kappa$ is the spring constant of the trap), and the Langevin
equations for the particle positions ($x_i,y_i$) are
\begin{equation}
\ddt{x_i} = - \frac{\partial U}{\partial x_i}
 + \gdot y_i + \eta_{x,i}\,,\quad
\ddt{y_i} =  - \frac{\partial U}{\partial y_i}
 + \eta_{y,i}\label{eq:Ntrap}
\end{equation}
where $\gdot$ is the shear rate and $\eta_{x,i}$ and $\eta_{y,i}$ are
noise terms defined as in the previous example.  We set $r_c=\kT=D=1$
to fix units of length, energy, and time, and choose $N=10$,
$\kappa=10$ and $\Gamma=25$; for this parameter set, the particles
form a dense, strongly correlated cluster in the trap (see inset to
Fig.~\ref{fig:trap25}). To characterize the morphology of the cluster
we use quantities like $\myav{xy} \equiv \myav{\sum_{i=1}^Nx_iy_i}/N$. We
first focus on the sensitivity of this quantity to changes in the
shear rate $\gdot$.  We therefore track the Malliavin weight
$q_{\gdot}$ which, from Eq.~\eqref{eq:gmw}, obeys
\begin{equation}
\begin{array}{l}
\displaystyle
\ddt{q_{\gdot}}=\frac{1}{2\kT}{\sum_{i=1}^N}
{y_i\eta_{x,i}}\\[6pt]
\end{array}
\label{eq:trapm}
\end{equation}
To compute $\partial\myav{xy}/\partial\gdot$ for the system in
steady-state, we use the correlation function approach outlined
above. Fig.~\ref{fig:trap25} shows that, for $\gdot=0$, $C(\Delta t) =
\myav{(1/N)(\sum_{i=1}^N x_i(t)y_i(t))
  (q_{\gdot}(t)-q_{\gdot}(t-\Delta t))}$ tends to
$\partial\myav{xy}/\partial\gdot|_{\gdot=0}$ as $\Delta t$ increases,
where $\partial\myav{xy}/\partial\gdot|_{\gdot=0}$
(the dashed line) is calculated by finite differencing.
In fact the Malliavin weight $q_{\gdot}$ turns out to be an
interesting physical quantity in itself. By splitting the sum in
Eq.~\eqref{eq:trapm} into individual particle contributions, one can
track Malliavin weights $q_{\gdot,i}$ for each individual
particle. These provide insight into the response to shear of the
one-particle probability distribution $P(x_i,y_i)$.  As shown in the
inset to Fig.~\ref{fig:trap25}b, the individual contributions to the
Malliavin weight are biased towards being positive in the first and
third quadrants and negative in the second and fourth quadrants,
corresponding to the distortion of the particle cloud by the shear.

\begin{figure}
\begin{center}
\includegraphics[clip=true,width=0.8\columnwidth]{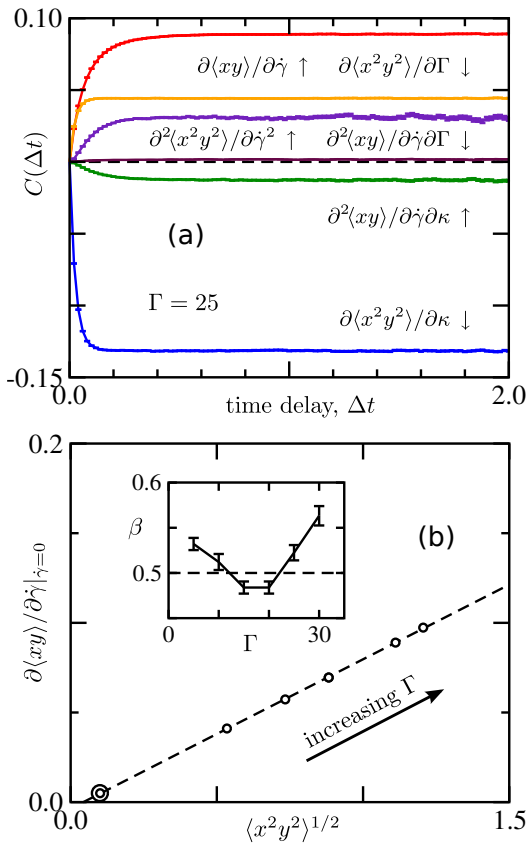}
\end{center}
\vskip -0.5cm
\caption{(color online) (a) Multiple parameter dependencies from a
  \emph{single} simulation.  (b) Dependence of shear response on the
  size of the cluster, as the interaction strength $\Gamma$ is varied
  (the double circle is $\Gamma=0$).  The inset shows the effective
  exponent $\beta$ computed from Eq.~\eqref{eq:beta} in the
  main text, using MWS to evaluate the derivatives.}
\label{fig:trapres}
\end{figure}

An important feature of MWS is that, from a single simulation run, one
can compute the sensivitities of \emph{any} function of the particle
coordinates, to \emph{any} parameter of the system. One simply needs to track
the Malliavin weights corresponding to all the parameters that are of
interest, using the appropriate dynamical rules as derived from
Eq.~\eqref{eq:gmw}. For example in this problem $q_\Gamma$ obeys
\begin{equation}
\begin{array}{l}
\displaystyle
\ddt{q_\Gamma}=-\frac{1}{2\kT}
{\sum_{i=1}^N}\Bigl(
\frac{\partial^2U}{\partial\Gamma\partial x_i}\eta_{x,i}
+\frac{\partial^2U}{\partial\Gamma\partial y_i}\eta_{y,i}
\Bigr)\,.
\end{array}
\label{eq:trapm2}
\end{equation}
and an analogous equation for $q_\kappa$ is easily written down.
Fig.~\ref{fig:trapres}a shows the full panoply of responses of the
cluster morphology parameters $\myav{xy}$ and $\myav{x^2y^2}$ to the
parameters of the problem, for $\Gamma=25$, obtained from a single
simulation run. Second order derivatives were computed as described in
Supplementary Material \cite{suppl-note}.

We now demonstrate how MWS can reveal subtle details of how the
response to shear, $\partial\myav{xy}/\partial\gdot|_{\gdot=0}$,
depends on $\myav{x^2y^2}^{1/2}$, which provides a representative
measure of the area of the particle cloud.  For non-interacting
particles one can obtain analytically \cite{suppl-note} the intriguing
quasi-FDT result $\partial\myav{xy}/\partial\gdot =
\myav{x^2y^2}/2T$, which holds at $\Gamma=\gdot=0$ and for all values
of $\kappa$.  The case of interacting particles, however, cannot be
solved analytically. We therefore simulated the cloud at a series of
increasing values of $\Gamma$ (\ie\, increasing cluster size) keeping
$\kappa=10$.  The results shown in Fig~\ref{fig:trapres}b suggest that
$\partial\myav{xy}/\partial\gdot|_{\gdot=0}$ is quite accurately
proportional to $\myav{x^2y^2}^{1/2}$.  More generally we can define
an effective exponent $\beta =
d\ln(\partial\myav{xy}/\partial\gdot) /
d\ln\myav{x^2y^2}$. This can be evaluated as $\Gamma$ varies at fixed
$\kappa$, or \viceversa.  In the former case, expanding the
derivatives gives
\begin{equation}
\beta=\frac{\myav{x^2y^2}}
{\partial\myav{x^2y^2}/\partial\Gamma}\times
\frac{\partial^2\myav{xy}/\partial\gdot\partial\Gamma}
{\partial\myav{xy}/\partial\gdot}\,.\label{eq:beta}
\end{equation}
We used MWS to compute the derivatives in Eq.~\eqref{eq:beta}, with
results shown in the inset in Fig.~\ref{fig:trapres}b.  As expected
from the main plot, $\beta\approx0.5$, but a more subtle dependence is
also apparent.  An analogous calculation for the latter case (fixed
$\Gamma$ varying $\kappa$) shows that the corresponding effective
exponent increases from $\beta=1$ at $\Gamma=0$ (the quasi-FDT result)
to $\beta\approx1.3$ at $\Gamma=25$.  Hence the interacting particle
case shows considerably more complexity than the non-interacting one,
and there is apparently little universality.

These examples demonstrate that MWS provides a simple and
easy-to-implement alternative to finite differencing, for Brownian
dynamics problems that may be time-dependent or in steady state, in or
out of equilibrium. Let us now discuss the question of efficiency. As
highlighted above, in MWS one has access to response coefficients for
all parameters of the problem (for which the derivatives $\partial
\fvec_i / \partial \lambda$ in Eq.~\eqref{eq:gmw} are known), with
little additional cost, since integrating the equations of motion for
the $q_\lambda$ requires no new random numbers and also typically does
not require recalculation of the forces. MWS also scales more
efficiently with the computational effort than does standard finite
differencing. For MWS, the error in a computation of $\partial
\myav{A(t)}/\partial \lambda$ scales as $M^{-1/2}$, where $M$ is the
number of replicate simulation runs used to compute the averages (this
follows from the usual scaling of the standard error in the mean with
the number of samples). For finite differencing, there is an inherent
tradeoff between the systematic error introduced by using a too-large
perturbation and the random sampling error that arises when the
perturbation is very small. One can show \cite{LE92} that the best
possible choice of the perturbation size results in an error that
scales as $M^{-1/4}$ for a forward finite differencing scheme, and
$M^{-1/3}$ for a centered scheme. Although the scaling can be improved
to $M^{-1/2}$ by using a common random number scheme \cite{LE92,
  RSK10}, even here we expect MWS to be more efficient, since it does
not require the perturbed system to be explicitly simulated.  For
example we note that more than six times as many force evaluations
were used to calculate $\partial\myav{xy}/\partial\gdot$ in
Fig.~\ref{fig:trap25} by finite differencing, compared to calculating
the same quantity to comparable accuracy by the MWS correlation
function method.

MWS has the potential greatly to facilitate the parameterization of
force fields, when combined with gradient-based search and
optimisation algorithms. This should be especially relevant for
mesoscale problems where Brownian dynamics algorithms such as
dissipative particle dynamics (DPD) are often the method of choice
\cite{FS02} (note that the application to DPD should take account of
the fact that the noise terms are pairwise central random forces).  An
equally important application of MWS is in the computation of response
functions to external fields---for example in the context of dynamical
phase transitions. Interestingly, while we have focused here only on
its long-time limit, the time-correlation function $C(\Delta t)$ is
actually equivalent to the time-dependent response to a step
perturbation. This point will be explored in more detail in future
work. A particularly interesting question concerns the extent to which
the Malliavin weight itself can be used as an autonomous order
parameter. For instance in the 1d trap problem the conditional average
$\myav{q_h}_x$ features the time-dependent effective temperature
$\Teff$, generalising the linear response result.  Certainly for
glassy systems we expect that the correlation functions
$\myav{A\,\Delta q_\lambda}$ will show typical non-ergodic memory and
aging phenomenology \cite{Ber07}: it is interesting to speculate
whether the behaviour of $q_\lambda$ itself could be used as an
alternative signature of glassy behaviour.

We thank Mike Allen, Mike Cates, Alessandro Laio and Bartlomiej Waclaw for
discussions. RJA is supported by a Royal Society University Research
Fellowship and by EPSRC under grants EP/I030298/1 and EP/EO30173.

%\bibliography{malliavin,malliavin_notes}

\begin{thebibliography}{17}
\bibitem[{alt()}]{alt-note}
\bibinfo{note}{As an alternative approach one can describe the Langevin
  dynamics as a hopping problem in $3N$-dimensional configuration space, using
  a master equation with a constant total hopping rate. Computation of
  sensitivity coefficients and response functions in master equation problems
  has been studied by Plyasunov and Arkin using the Girsanov measure transform
  \cite{PA07}, by ourselves using the notion of trajectory weights \cite{WA12};
  and by Berthier in the context of Monte-Carlo simulations \cite{Ber07}.}
\end{thebibliography}

%%%% START OF SUPPLEMENTARY MATERIAL

\renewcommand{\thefigure}{S\arabic{figure}}
\renewcommand{\theequation}{S\arabic{equation}}

\setcounter{section}{0}
\renewcommand{\thesection}{S\arabic{section}}

\setcounter{equation}{0}
\setcounter{figure}{0}

\begin{widetext}
\begin{center}
\vspace{9pt}
\framebox{\bf\Large SUPPLEMENTARY MATERIAL}
\end{center}
\end{widetext}

%\large

\mysection{Proof of Eqs.~(3) and~(4) in the main text}
\label{sec:eq4}
We start by introducing the probability distribution $P(\{\rvec_i\};
t)$ for the particle positions, and the joint probability distribution
$P(\{\rvec_i\}, q_\lambda; t)$ for the combination of the particle
positions and the Malliavin weight.  The two are related by
\begin{equation}
\textstyle
P(\{\rvec_i\};t)
=\int\! dq_\lambda\,P(\{\rvec_i\},q_\lambda;t)\,.
\label{supp:eq:pdef}
\end{equation}
We also define the conjugate variable
\begin{equation}
Q_\lambda(\{\rvec_i\};t) \equiv
\frac{\partial\ln P(\{\rvec_i\};t)}{\partial\lambda}
=
\frac{1}{P(\{\rvec_i\};t)}
\frac{\partial P(\{\rvec_i\};t)}{\partial\lambda}
\,,
\label{supp:eq:conj}
\end{equation}
and the conditional average
\begin{equation}
\myav{q_\lambda}_{\{\rvec_i\}}\equiv
\frac{\int\!dq_\lambda\,q_\lambda\,P(\{\rvec_i\},q_\lambda;t)}
{\int\!dq_\lambda\,P(\{\rvec_i\},q_\lambda;t)}\,.
\label{supp:eq:cond}
\end{equation}
The first goal is to show the equivalence of $Q_\lambda$ and
$\myav{q_\lambda}_{\{\rvec_i\}}$---Eq.~\eqref{eq:condav} in the main
text.  We do this by showing that they both obey the same evolution
equation.

As hinted at in the main text, we introduce an explicit Euler-type
scheme for updating the particle positions.  This also makes explicit
the updating rule associated with the Malliavin weight.  We therefore
write
\begin{equation}
\rvec_i{}'=\rvec_i + \frac{D\fvec_i
\dt}{\kT} + \xivec_i
\label{supp:eq:rupdate}
\end{equation}
where $\rvec_i{}'$ is the updated position of the $i$th particle at
time $t+\dt$, $\dt$ is the time step, and the $\xivec_i$ are a set of
$3N$ Gaussian random variates of zero mean and variance $2D\dt$.  The
corresponding updating rule for the Malliavin weight is
\begin{equation}
q_\lambda'=q_\lambda + \frac{1}{2\kT}\sum_{i=1}^N
\frac{\partial\fvec_i}{\partial\lambda}\cdot\xivec_i\,.
\label{supp:eq:qupdate}
\end{equation}
Note that the exact same sequence of random variates $\xivec_i$ is
used for updating the particle positions and the Malliavin weight.  In
this Euler scheme $P(\{\rvec_i\};t)$ obeys a Chapman-Kolmogorov
equation
\begin{equation}
\begin{array}{l}
P(\{\rvec_i{}'\};t+\dt)={\textstyle\int \prod_i d^3\rvec_i}\,
P(\{\rvec_i\};t)\\[6pt]
{}\hspace{12em}\,\times W(\{\rvec_i{}'\}|\{\rvec_i\})
\end{array}
\end{equation}
where the propagator is 
\begin{widetext}
\begin{equation}
W(\{\rvec_i{}'\}|\{\rvec_i\})=
{({4\pi D\,\dt})^{-3N/2}}
\exp\Bigl(-\frac{\sum_{i=1}^N(\rvec_i{}'-\rvec_i-D\fvec_i\dt/\kT)^2}
{4 D\,\dt}\Bigr)\,.
\label{supp:eq:emprop}
\end{equation}
Differentiating the Chapman-Kolmogorov equation with respect to
$\lambda$ leads to an adjunct equation for the conjugate variable
$Q_\lambda$,
\begin{equation}
%\begin{array}{l}
Q_\lambda(\{\rvec_i{}'\};t+\dt)\,P(\{\rvec_i{}'\};t+\dt)
=\int \prod_i d^3\rvec_i\, 
\Bigl[Q_\lambda(\{\rvec_i\};t)
+\frac{\partial\ln W}{\partial\lambda}\Bigr]%\\[6pt]
\,%{}\hspace{21em}\times 
P(\{\rvec_i\};t)\,W(\{\rvec_i{}'\}|\{\rvec_i\})\,.
%\end{array}
\label{supp:eq:ack}
\end{equation}
The second quantity in the square brackets in Eq.~\eqref{supp:eq:ack} is
\begin{equation}
\frac{\partial\ln W}{\partial\lambda}
\equiv\frac{\partial\ln W(\{\rvec_i{}'\}|\{\rvec_i\})}{\partial\lambda}
=\frac{1}{2\kT}\sum_{i=1}^N
\frac{\partial\fvec_i}{\partial\lambda}\cdot
\Bigl(\rvec_i{}'-\rvec_i-\frac{D\fvec_i\dt}{\kT}\Bigr)\,.
\label{supp:eq:lnw}
\end{equation}

We now show that Eq.~\eqref{supp:eq:ack} for the evolution of $Q_\lambda$
is replicated by the evolution equation for the Malliavin weight.  The
joint probability distribution function $P(\{\rvec_i\}, q_\lambda; t)$
obeys an extended Chapman-Kolmogorov equation
\begin{equation}
%\begin{array}{l}
P(\{\rvec_i{}'\},q_\lambda';t+\dt)=
\int \prod_i d^3\rvec_i\,d^3\xivec_i\, \,dq_\lambda\,
P(\{\rvec_i\},q_\lambda;t)\,
P(\{\xivec_i\})\,%\\[6pt]
%{}\hspace{14em}\times
\delta(\rvec_i{}'-\rvec_i-\delr)\,
\delta(q_\lambda'-q_\lambda-\delq)
%\end{array}
\label{supp:eq:ck2}
\end{equation}
where
\begin{equation}
\delr(\rvec_i,\xivec_i)=\frac{D\fvec_i\dt}{\kT}+\xivec_i\,,\qquad
\delq(\rvec_i,\xivec_i)=\frac{1}{2\kT}\sum_{i=1}^N
\frac{\partial\fvec_i}{\partial\lambda}\cdot\xivec_i
\end{equation}
give the discrete increments to the particle positions and Malliavin
weight.  The two $\delta$-functions in Eq.~\eqref{supp:eq:ck2} enforce
these updating rules.  The function $P(\{\xivec_i\})$ is a
$3N$-dimensional Gaussian.  The $3N$ random variates $\xivec_i$ are
uncorrelated and have zero mean and variance $2D\dt$.  With the
definitions in Eqs.~\eqref{supp:eq:conj} and \eqref{supp:eq:cond} in hand we take
the first moment of Eq.~\eqref{supp:eq:ck2} with respect to $q_\lambda$ to
get
\begin{subequations}
\begin{align}
&\textstyle\myav{q_\lambda}_{\{\rvec_i{}'\},t+\dt}\,P(\{\rvec_i{}'\};t+\dt)
=\int\!dq_\lambda'\,q_\lambda'\,P(\{\rvec_i{}'\},q_\lambda';t+\dt)\\[6pt]
&\textstyle{}\quad=
\int \prod_i d^3\rvec_i\,d^3\xivec_i\,dq_\lambda\,dq_\lambda'\, 
q_\lambda'\,
P(\{\rvec_i\},q_\lambda;t)\,P(\{\xivec_i\})\,\delta(\rvec_i{}'-\rvec_i-\delr)\,
\delta(q_\lambda'-q_\lambda-\delq)\label{supp:eq:mid}\\[6pt]
&\textstyle{}\quad=
\int \prod_i d^3\rvec_i\,d^3\xivec_i\,dq_\lambda\,
[q_\lambda+\delq]\,
P(\{\rvec_i\},q_\lambda;t)\,P(\{\xivec_i\})\,
\delta(\rvec_i{}'-\rvec_i-\delr)\\[6pt]
&\textstyle{}\quad=
\int \prod_i d^3\rvec_i\,dq_\lambda\,
[q_\lambda+\delq]\,
P(\{\rvec_i\},q_\lambda;t)\,W(\{\rvec_i{}'\}|\{\rvec_i\})\\[6pt]
&\textstyle{}\quad=
\int \prod_i d^3\rvec_i\,
[\myav{q_\lambda}_{\{\rvec_i\}}+\delq]\,
P(\{\rvec_i\};t)\,W(\{\rvec_i{}'\}|\{\rvec_i\})\,.
\label{supp:eq:end}
\end{align}
\end{subequations}
To progress from Eq.~\eqref{supp:eq:mid} to \eqref{supp:eq:end} we do
successively the $q_\lambda'$ integral, the $\xivec_i$ integrals, and
the $q_\lambda$ integral; using first the $\delta$-functions then the
definition of the conditional average for the final step.  In doing
the $\xivec_i$ integrations, we recover the propagator
$W(\rvec_i{}'|\rvec_i)$ given by Eq.~\eqref{supp:eq:emprop} and the
$q_\lambda$ increment becomes
\begin{equation}
\delq(\rvec_i,\rvec_i{}')=\frac{1}{2\kT}\sum_{i=1}^N
\frac{\partial\fvec_i}{\partial\lambda}\cdot
\Bigl(\rvec_i{}'-\rvec_i-\frac{D\fvec_i\dt}{\kT}\Bigr)\,.
\end{equation}
This is identical to Eq.~\eqref{supp:eq:lnw} therefore we conclude that
Eq.~\eqref{supp:eq:end} for updating $\myav{q_\lambda}_{\rvec_i}$ is
identical to Eq.~\eqref{supp:eq:ack} for updating the conjugate variable
$Q_\lambda$.  By choice these two quantities can be given the same
initial values, thus establishing their equivalence, in other words
\begin{equation}
\frac{\int\!dq_\lambda\,q_\lambda\,P(\{\rvec_i\},q_\lambda;t)}
{\int\!dq_\lambda\,P(\{\rvec_i\},q_\lambda;t)} = 
\frac{\partial \ln P(\{\rvec_i\};t)}{\partial \lambda}\,.
\label{supp:eq:condav}
\end{equation}
To establish  Eq.~\eqref{eq:mall} in the main text we use
Eq.~\eqref{supp:eq:pdef} and the second half of Eq.~\eqref{supp:eq:conj} to
rearrange Eq.~\eqref{supp:eq:condav} into the alternative form
$\int\!dq_\lambda\,q_\lambda\,P(\{\rvec_i\},q_\lambda;t) ={\partial
  P(\{\rvec_i\};t)}/{\partial \lambda}$.  Eq.~\eqref{eq:mall}
follows from this since
\begin{equation}
\begin{array}{l}
\myav{A q_\lambda}=\int \prod_i d^3\rvec_i\,dq_\lambda\,
A(\{\rvec_i\})\,q_\lambda\,P(\{\rvec_i\},q_\lambda;t)%\\[6pt]
%\displaystyle{}\hspace{6em}
\\[12pt]
\displaystyle{}\hspace{4em}
={\textstyle \int \prod_i d^3\rvec_i}\,A(\{\rvec_i\})\,
\frac{\partial P(\{\rvec_i\};t)}{\partial\lambda}
=\frac{\partial({\textstyle\int \prod_i d^3\rvec_i}\,A(\{\rvec_i\})\,
P(\{\rvec_i\};t))}{\partial\lambda}
=\frac{\partial\myav{A}}{\partial\lambda}\,.
\end{array}
\end{equation}
\end{widetext}

\mysection{Practical implementation}
\label{sec:inotes}
The implementation of MWS in a Brownian dynamics code is quite
straightforward.  The main point is that for efficiency one may want
to update the Malliavin weight(s) according to Eq.~\eqref{supp:eq:qupdate}
at the same time as calculating the forces.  This requires that the
set of random variates $\xivec_i$ be computed and stored at the start
of the step, as in the following schematic algorithm :
\begin{itemize}
\item generate the $3N$ random variates $\xivec_i$,
\item compute the forces $\fvec_i(\{\rvec_i\})$ and the quantity 
$\sum_{i=1}^N(\partial\fvec_i/\partial\lambda)\cdot\xivec_i$,
\item update the particle positions according in Eq.~\eqref{supp:eq:rupdate},
\item update the Malliavin weight(s) according in Eq.~\eqref{supp:eq:qupdate},
\item record any quantities of interest, \ie\
$A(\{\rvec_i\})$, and the value of the
  Malliavin weight(s).
\end{itemize}
The last item does not necessarily have to be done every time step of
course.  As indicated in the main text, the algorithm is initialised
by setting the particle positions to their initial values and the
Malliavin weights to zero.  For steady state problems there are two
approaches to the use of the MWS correlation function method.  The
simplest is to choose a set of equally spaced reference points
$t_0=nT$ for the calculation of $\Delta q_\lambda = q_\lambda(t) -
q_\lambda(t_0)$, where $T$ is a time period longer than the expected
relaxation time of the system (which may have to be determined by
trial and error).  In this approach, every $n$ timesteps the current
values of $q_\lambda(t)$ and $A(t)$ are recorded. The running value of
$q_\lambda$ is then set to zero and the simulation continued. The time
average of the stored values of $q_\lambda A$ gives $\partial
\myav{A}/\partial \lambda$.  More efficient is to use a sliding window
to calculate the correlation function.  Block averaging can be used
for error estimates.

\begin{figure}
\begin{center}
\includegraphics[clip=true,width=0.8\columnwidth]{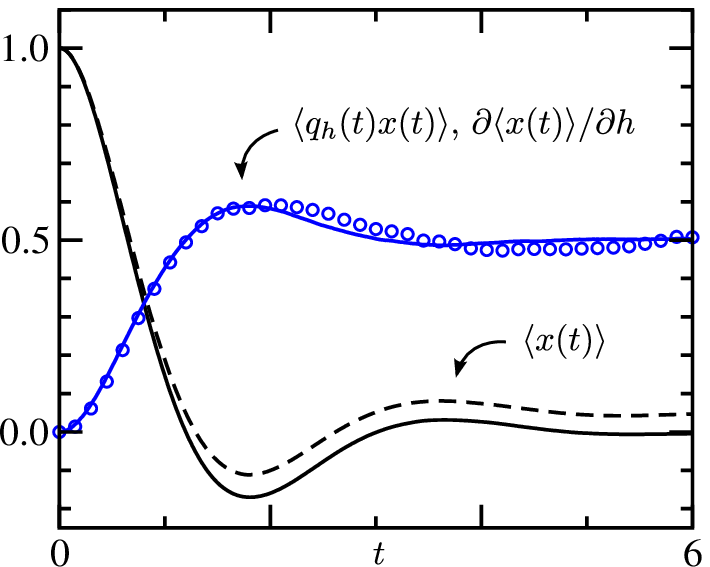}
\end{center}
\vskip -0.5cm
\caption{(color) Numerical simulation of Eqs.~\eqref{supp:eq:e1trap1}
  and ~\eqref{supp:eq:e1trap2} with $\kappa = 2$, $m=1/2$, $x_0=1$,
  and $\gamma=T=1$.  The falling curves show $\myav{x(t)}$ for $h=0$
  (solid line) and $h=0.1$ (dashed line).  The rising curves show
  $\myav{x(t)q_h(t)}$ (blue solid line) and
  $\partial\myav{x(t)}/\partial h$ evaluated using forward finite
  differencing (blue circles). Averages are over $10^5$ replicate
  simulations; \cf\ Fig.~\ref{fig:trap1d} in the main text.}
\label{fig:etrap1d}
\end{figure}

\mysection{Underdamped Brownian dynamics}
\label{sec:underdamp}
The extension of MWS to underdamped Brownian dynamics is fairly
simple.  We denote the particle positions by $\rvec_i$ and velocities
by $\vvec_i$.  The dynamical equations are
\begin{equation}
\ddt{\rvec_i}=\vvec_i\,,\quad
m\ddt{\vvec_i}=\fvec_i-\gamma\vvec_i+\etavec_i
\label{supp:eq:ubd}
\end{equation}
where $m$ is the mass, $\gamma$ is the frictional drag coefficient,
and $\etavec_i$ are white noise terms of amplitude $2\gamma\kT$.  
An Euler-type scheme for Eqs.~\eqref{supp:eq:ubd} is
\begin{equation}
\rvec_i{}'=\rvec_i+\vvec_i\dt\,\,\quad
\vvec_i{}'=\vvec_i+\frac{(\fvec_i-\gamma\vvec_i)\dt}{m}+\xivec_i
\end{equation}
where the $\xivec_i$ are 3N Gaussian random variates of zero mean and
variance $2\gamma\kT\,\dt/m^2$ (note that we divided the velocity
equations through by $m$).  It follows that the probability
distribution function $P(\{\rvec_i\},\{\vvec_i\};t)$ evolves according to the
Chapman-Kolmogorov equation
\begin{widetext}
\begin{equation}
%\begin{array}{l}
P(\{\rvec_i{}'\},\{\vvec_i{}'\};t+\dt)=
\int \prod_i d^3\rvec_i\,d^3\vvec_i\,
P(\{\rvec_i\},\{\vvec_i\};t)%\\[6pt]
%{}\hspace{12em}\times
\,
\delta(\rvec_i{}'-\rvec_i-\vvec_i\dt)\,
W(\{\vvec_i{}'\}|\{\vvec_i\},\{\rvec_i\})
%\end{array}
\label{supp:eq:eprop}
\end{equation}
where the partial propagator is
\begin{equation}
W(\{\vvec_i{}'\}|\{\vvec_i\},\rvec_i)=
(4\pi\gamma\kT\,\dt/m^2)^{-3N/2}
\exp\Bigl(-\frac{\sum_{i=1}^N
(\vvec_i{}'-\vvec_i-(\fvec_i-\gamma\vvec_i)\dt/m)^2}
{4 \gamma\kT\,\dt/m^2}\Bigr)\,.
\end{equation}
If we differentiate Eq.~\eqref{supp:eq:eprop} with respect to some
parameter $\lambda$ we obtain 
\begin{equation}
\begin{array}{l}
Q_\lambda(\{\rvec_i{}'\},\{\vvec_i{}'\};t+\dt)\,
P(\{\rvec_i{}'\},\{\vvec_i{}'\};t+\dt)\\[6pt]
{}\hspace{6em}=\int \prod_i d^3\rvec_i\,d^3\vvec_i 
[Q_\lambda(\{\rvec_i\},\{\vvec_i\};t)+{\partial\ln W}/{\partial\lambda}]\\[6pt]
{}\hspace{12em} \times 
\delta(\rvec_i{}'-\rvec_i-\vvec_i\dt)\,
W(\{\vvec_i{}'\}|\{\vvec_i\},\{\rvec_i{}'\})\,
P(\{\rvec_i\},\{\vvec_i\};t)
\end{array}
\label{supp:eq:eack}
\end{equation}
\end{widetext}
where $Q_\lambda(\{\rvec_i\},\{\vvec_i\}; t)\equiv \partial\ln
P(\{\rvec_i\},\{\vvec_i\}; t) / \partial\lambda$.  This again suggests
the updating rule for the Malliavin weight, $q_\lambda' = q_\lambda +
\partial\ln W/\partial\lambda$, in other words the derivation is
impervious to the presence of a $\delta$-function in the full
propagator.  Inserting the explicit expression for $W$, and assuming
the parameter of interest features only in the force law, gives the
Langevin equation
\begin{equation}
\ddt{q_\lambda}=\frac{m}{2\gamma\kT} \sum_{i=1}^N 
\frac{\partial\fvec_i}{\partial\lambda}\cdot\etavec_i\,.
\label{supp:eq:egmw}
\end{equation}
This result is very similar to the overdamped case.
As a demonstration let us revisit the example of Brownian
motion in a one-dimensional trap, but this time consider the
underdamped case.  The Langevin equations are
\begin{equation}
\ddt{x}=v,\quad
m\ddt{v}=-\gamma v-\kappa x+ h+\eta,
\label{supp:eq:e1trap1}
\end{equation}
and
\begin{equation}
\ddt{q_h}=\frac{m\eta}{2\gamma T}\,.
\label{supp:eq:e1trap2}
\end{equation}
Fig.~\ref{fig:etrap1d} shows simulation results confirming that
$\myav{q_h x}=\partial\myav{x}/\partial h$.  

\mysection{Higher-order derivatives}
\label{sec:horder}
We next demonstrate how MWS extends to the computation of higher-order
derivatives.  A double application of Eq.~\eqref{eq:mall} in the
main text, for two parameters $\lambda$ and $\mu$, gives
\begin{equation}
\frac{\partial^2\!\myav{A(\{\rvec_i\})}}{\partial\lambda\partial\mu}
=\myav{A(\{\rvec_i\})\,(q_{\lambda\mu}+q_\lambda q_\mu)}
\label{supp:eq:mal2}
\end{equation}
where $q_{\lambda\mu}=\partial q_\lambda/\partial\mu$.  
Differentiating the discrete updating rule
$q_\lambda'=q_\lambda+\partial\ln W/\partial\lambda$ with respect to
$\mu$ gives $q_{\lambda\mu}'= q_{\lambda\mu}+\partial^2\!\ln
W/\partial\lambda\partial\mu$.  We insert the expression for $W$ from
Eq.~\eqref{supp:eq:emprop} into this, and simplify, to
find the corresponding Langevin equation
\begin{equation}
\ddt{q_{\lambda\mu}}=
\frac{1}{2\kT} \sum_{i=1}^N \Bigl[
\frac{\partial^2\!\fvec_i}{\partial\lambda\partial\mu}\cdot\etavec_i
-\frac{D}{\kT}
\frac{\partial\fvec_i}{\partial\lambda}
\cdot\frac{\partial\fvec_i}{\partial\mu}\Bigr]\,.
\label{supp:eq:gmw2}
\end{equation}
\vspace{0pt} % otherwise we have too little space after the eq for
             % some reason...

\noindent 
The new feature in this is a drift term (the last term) which has the
consequence that $\myav{q_{\lambda\mu}}=-\myav{q_\lambda q_\mu}$.
This ensures that ${\partial^2\!\myav{A}} /
{\partial\lambda\partial\mu} = 0$ if $A$ is a constant.

Note that, as a result of the peculiar properties of the stochastic
differential calculus, Eq.~\eqref{supp:eq:gmw2} is not simply found by
differentiating Eq.~\eqref{eq:gmw} in the main text; rather one
has to proceed \via\ the discrete updating rules.  Also note that the
second-order Malliavin weight $q_{\lambda\mu}$ defined in
Eq.~\eqref{supp:eq:gmw2} must be combined with the two first-order
Malliavin weights $q_\lambda$ and $q_\mu$ to obtain the correct
weighted average in Eq.~\eqref{supp:eq:mal2}.  We have tested
the second-order MWS scheme for the trapped interacting particle cloud
under shear, see for example Fig~\ref{fig:trapres}a in the main
text.

\mysection{One-dimensional trap}
\label{sec:1dtrap}
Here we derive the expressions in Eqs.~\eqref{eq:1dtrap} for the
transient behaviour of a particle in a one-dimensional harmonic trap.
Eqs.~\eqref{eq:1trap} in the main text are
\begin{equation}
\ddt{x}=-\kappa x+ h+\eta,\quad
\ddt{q_h}=\frac{\eta}{2T}\,.\label{supp:eq:1trap}
\end{equation}
These can be integrated to find
\begin{equation}
\begin{array}{l}
\displaystyle
x(t)=x_0 e^{-\kappa t} + \frac{h}{\kappa}(1-e^{-\kappa t})
+\int_0^t \!dt'\,e^{-\kappa(t-t')}\eta(t')\,,\\[12pt]
\displaystyle
q_h(t)=\frac{1}{2T}\int_0^t \!dt'\,\eta(t')\,.
\end{array}
\label{supp:eq:1tint}
\end{equation}
Since $x$ and $q_h$ are summed Gaussian random noises, it
follows that $P(x,q_h;t)$ is a Gaussian---\cf\ \S 3.5.2 in \emph{The
  Theory of Polymer Dynamics} M.~Doi and S.~F.~Edwards (OUP, 1986).
To characterise this Gaussian, it suffices to calculate the first and
second moments.  The first moments follow immediately from
Eqs.~\eqref{supp:eq:1tint} and are $\myav{x}=x_0 e^{-\kappa t} +
({h}/{\kappa})(1-e^{-\kappa t})$ and $\myav{q_h}=0$, as given in the
first line of Eqs.~\eqref{eq:1dtrap}.  Therefore
$x-\myav{x}=\int_0^t \!dt'\,e^{-\kappa(t-t')}\eta(t')$. The second moment 
of $x$ (the second line of Eqs.~\eqref{eq:1dtrap}) is 
\begin{widetext}
\begin{equation}
\myav{x^2}-\myav{x}^2=\int_0^t\!dt'\!\int_0^t\!dt''\,
e^{-\kappa(t-t')}\,e^{-\kappa(t-t'')}\times 2T\,\delta(t'-t'')
= \frac{T}{\kappa}(1-e^{-2\kappa t})
\end{equation}
where we have used $\myav{\eta(t')\eta(t'')}=2T\,\delta(t'-t'')$.
Likewise the cross correlation term and the second moment of $q_h$
(the third line of Eqs.~\eqref{eq:1dtrap}) are
\begin{equation}
\begin{array}{l}
\displaystyle
\myav{x q_h}
=\frac{1}{2T}\int_0^t\!dt'\!\int_0^t\!dt''\,
e^{-\kappa(t-t')}\times 2T\,\delta(t'-t'')
= \frac{1}{\kappa}(1-e^{-\kappa t})\\[12pt]
\displaystyle
\myav{q_h^2}
=\frac{1}{(2T)^2}\int_0^t\!dt'\!\int_0^t\!dt''\,
\times 2T\,\delta(t'-t'')
= \frac{t}{2T}
\end{array}
\end{equation}
\end{widetext}

\mysection{Two-dimensional trap in shear}
\label{sec:2dtrap}
The quasi-FDT result in the main text follows from the steady state
probability distribution $\Pss(x,y)$ for a particle in a
two-dimensional trap under shear, which can be solved in closed form.
Let us recall the Langevin equations for this problem,
\begin{equation}
\ddt{x}=-\kappa x+\gdot y+\eta_x,\quad
\ddt{y}=-\kappa y+\eta_y\,,
\end{equation}
where $\kappa$ is the trap strength and $\gdot$ is the shear rate.  In
common with the main text we set the particle mobility to unity and
write temperature in terms of Boltzmann's constant, so that we can
write $D=T$ for the diffusion coefficient.

From the Smoluchowski equation, the steady-state distribution function
corresponding to these Langevin equations satisfies
\begin{equation}
\frac{\partial J_x}{\partial x}
+\frac{\partial J_y}{\partial y}=0\label{supp:eq:s}
\end{equation}
where the components of the probability current (flux) are
\begin{equation}\begin{array}{l}
\displaystyle
J_x=(-\kappa x + \gdot y) \Pss
- T \frac{\partial\Pss}{\partial x},\\[12pt]
\displaystyle
J_y= -\kappa y \Pss
- T \frac{\partial\Pss}{\partial y}\,.
\end{array}
\label{supp:eq:j}
\end{equation}
Since the Langevin equations are linear, we expect that $\Pss$ will be
a bivariate Gaussian so we write
\begin{equation}
\textstyle\Pss \sim\exp( -\frac{1}{2}A x^2-\frac{1}{2}B y^2-C xy)
\end{equation}
where $A$, $B$ and $C$ are coefficients, to be determined.  
The simplest way to proceed is to insert this as an \ansatz\ into
Eqs.~\eqref{supp:eq:s} and~\eqref{supp:eq:j}, to find that the coefficients have
to satisfy
\begin{equation}
\begin{array}{l}
2\kappa = (A + B)T,\\[3pt]
A\kappa = (A^2 + C^2)T,\\[3pt]
B\kappa = C\gdot + (B^2 + C^2)T,\\[3pt]
C\kappa = A\gdot/2 + (A + B)CT.
\end{array}
\label{supp:eq:lin}
\end{equation}
These four conditions arise from equating to zero the constant term
and the coefficients of $x^2$, $y^2$ and $xy$ in Eq.~\eqref{supp:eq:s}.
Although there are only three unknowns, Eqs.~\eqref{supp:eq:lin} are
interdependent and admit the unique solution,
\begin{equation}
\begin{array}{l}
\displaystyle
A=\frac{4 \kappa ^3}{({\gdot}^2+4 \kappa^2)T}\,,\quad
B=\frac{4 \kappa ^3+2 \kappa\gdot^2}
{({\gdot}^2+4 \kappa ^2)T}\,,\\[12pt]
\displaystyle
C=-\frac{2 {\gdot} \kappa ^2}{({\gdot}^2+4 \kappa ^2)T}\,.
\end{array}
\end{equation}
Hence the complete steady state distribution function is
\begin{equation}
\begin{array}{l}
\displaystyle
\Pss=\frac{\kappa ^2}{\pi  T \sqrt{\gdot^2+4 \kappa ^2}}\\[12pt]
\displaystyle{}\hspace{1em} \times 
\exp\Bigl[
-\frac{2\kappa^3 x^2+(2\kappa^3+\kappa\gdot^2)y^2-2 \gdot \kappa^2  x
  y}
{(\gdot^2+4 \kappa ^2)T}
\Bigr]\,.
\end{array}
\end{equation}
For reference, the associated moments are
\begin{equation}
\begin{array}{l}
\displaystyle
\myav{x^2}_{\mathrm{ss}}=\frac{T}{\kappa}\Bigl(1 + 
\frac{\gdot^2}{2\kappa^2}\Bigr)\,,\quad
\myav{y^2}_{\mathrm{ss}}=\frac{T}{\kappa}\,,\\[12pt]
\displaystyle
\myav{xy}_{\mathrm{ss}}=\frac{\gdot T}{2\kappa^2}\,.
\end{array}
\end{equation}
Differentiating the steady state distribution with respect to the
shear rate yields
\begin{equation}
\begin{array}{l}
\displaystyle
\myav{q_{\gdot}}_{xy}=\frac{\partial\ln\Pss}{\partial\gdot} \\[12pt]
\displaystyle{}\hspace{2em}=\frac{2 \kappa ^2 (2 \kappa
  x-\gdot y) (\gdot x+2 \kappa y)} { (\gdot^2+4 \kappa
  ^2)^2T}-\frac{\gdot}{\gdot^2+4 \kappa ^2}\,.
\end{array}
\end{equation}
In the limit $\gdot\to0$ this reduces to $\myav{q_{\gdot}}_{xy} = xy /
2T$.  An immediate application of this is to deduce the quasi-FDT
result in the main text:
\begin{equation}
\frac{\partial\myav{xy}}{\partial\gdot} \Big|_{\gdot=0}
%=\myav{xyq_{\gdot}}=\myav{xy\myav{q_{\gdot}}_{xy}}
=\frac{\myav{x^2y^2}}{2T}\,.
\end{equation}

\end{document}